\documentclass{article}
\usepackage{spconf,amsmath,graphicx,hyperref}

\title{Bridging the Inter-Domain Gap through Low-Level Features for Cross-Modal Medical Image Segmentation}
\usepackage{cite}
\usepackage{amsmath,amssymb,amsfonts}
\usepackage{algorithmic}
\usepackage{graphicx}
\usepackage{textcomp}
\usepackage{xcolor}

\newcommand*{\eg}{\emph{e.g.}\xspace}
\newcommand*{\ie}{\emph{i.e.}\xspace}
\newcommand*{\proposed}{LowBridge\xspace}
\usepackage{xspace}
\usepackage{url}
\usepackage{hyperref}
\hypersetup{hidelinks, colorlinks=True, allcolors=magenta, pdfstartview=Fit, breaklinks=true}
\usepackage{array}
\usepackage{booktabs}
\usepackage{multirow}

\begin{document}
\ninept


\name{
\begin{tabular}{c}
Pengfei Lyu$^{1,\dag}$\thanks{$\dag$ These authors contributed equally.} \qquad Pak‑Hei Yeung$^{2,\dag}$ \qquad Jing Xia$^{3,*}$\thanks{$*$ Corresponding author.} \qquad De Hu$^{1}$ \qquad Xiaosheng Yu$^{4}$\\
Jianning Chi$^{4}$ \qquad Chengdong Wu$^{4}$ \qquad Jagath C. Rajapakse$^{5,*}$
\end{tabular}
}

\address{$^1$ College of Computer Science, Inner Mongolia University, Hohhot, China \\
      $^2$ Informatics Institute, University of Amsterdam, Amsterdam, Netherlands \\
      $^3$ College of Biomedical Engineering and Instrument Science, Zhejiang University, Hangzhou, China \\
      $^4$ Faculty of Robot Science and Engineering, Northeastern University, Shenyang, China\\
      $^5$ College of Computing and Data Science, Nanyang Technological University, Singapore}
\maketitle
\begin{abstract}
This paper addresses cross-modal medical image segmentation, focusing on MRI-CT transfer in a \textit{source-only} domain generalization setting.
During training, only source-modality samples are available, while unlabeled target-modality images are used for testing.
We propose \proposed, which builds on the observation that cross-modal images share similar low-level features (\eg edges) as they depict the same types of anatomical structures. 
Specifically, we first train a generative model to recover the source images from their edge features, followed by training a segmentation model on the generated source images, separately.
At test time, edge features from the target images are input to the pretrained generative model to generate source-style target domain images, which are then segmented using the pretrained segmentation network. 
Experiments on various public datasets demonstrate that \proposed achieves state-of-the-art performance, outperforming ten existing approaches. Ablation studies further show that \proposed is compatible with different types of generative and segmentation models, suggesting its generalizability and potential to benefit from future advances in these models. The code will be available at \url{https://github.com/JoshuaLPF/LowBridge}.
\end{abstract}

\begin{keywords}
Domain Generalization, Cross-Modal Image Segmentation, Low-Level Features, MRI, CT
\end{keywords}

\section{Introduction}
\label{sec:intro}
Image segmentation is an essential step in medical image analysis, 
serving as the foundation for various downstream tasks such as diagnosis \cite{zhou2020rapid} and treatment planning \cite{gonzalez2021semi}. 
Recent advances in deep learning have greatly improved the accuracy of automatic segmentation \cite{ren2025hresformer,xing2024segmamba}, 
especially when the training and test data belong to the same domain (\ie modality). 
However, when tested on data from different domains, deep learning models often suffer significant performance degradation due to domain shift.
Although incorporating labeled data from the target domain could help mitigate this issue, acquiring large-scale annotations across different domains is often impractical in clinical practice because of the high cost and substantial effort required for manual annotation\cite{zhang2018translating}.
This limitation is particularly pronounced in cross-modal medical image segmentation,
where labeled images are available for one modality (\eg MRI) but not for another (\eg CT).

To address this challenge, existing approaches have primarily adopted the unsupervised domain adaptation (UDA) paradigm, which transfers knowledge from a labeled source domain to an unlabeled target domain.
Recent UDA methods have achieved impressive results, by aligning distributions in feature space \cite{Vu_2019_CVPR,pei2021disentangle,ji2023unsupervised,pei2023multi,feng2023unsupervised}, 
enhancing feature discriminability through self-paced contrastive objectives \cite{liu2022margin}, 
and synthesizing target-style images from the source data \cite{chen2019synergistic,chen2020unsupervised}.
Despite the remarkable performance achieved, 
alignment in the feature space may fail to capture domain-specific characteristics, 
particularly when modalities differ drastically in appearance.
Also, synthetic images often struggle to accurately represent the target domain due to their insensitivity to geometric details or lack of semantic consistency. 
Moreover, some methods \cite{chen2019synergistic,pei2021disentangle, liu2022margin,feng2023unsupervised,pei2023multi} rely on complex training objectives, 
such as generative adversarial networks, 
which are difficult to train and require substantial computational resources. More importantly, these UDA methods use unlabeled target‑modality samples during training, which may not be available in real-world clinical deployments.

\begin{figure*}[!htp]
	\centering \includegraphics[width=0.88\textwidth]{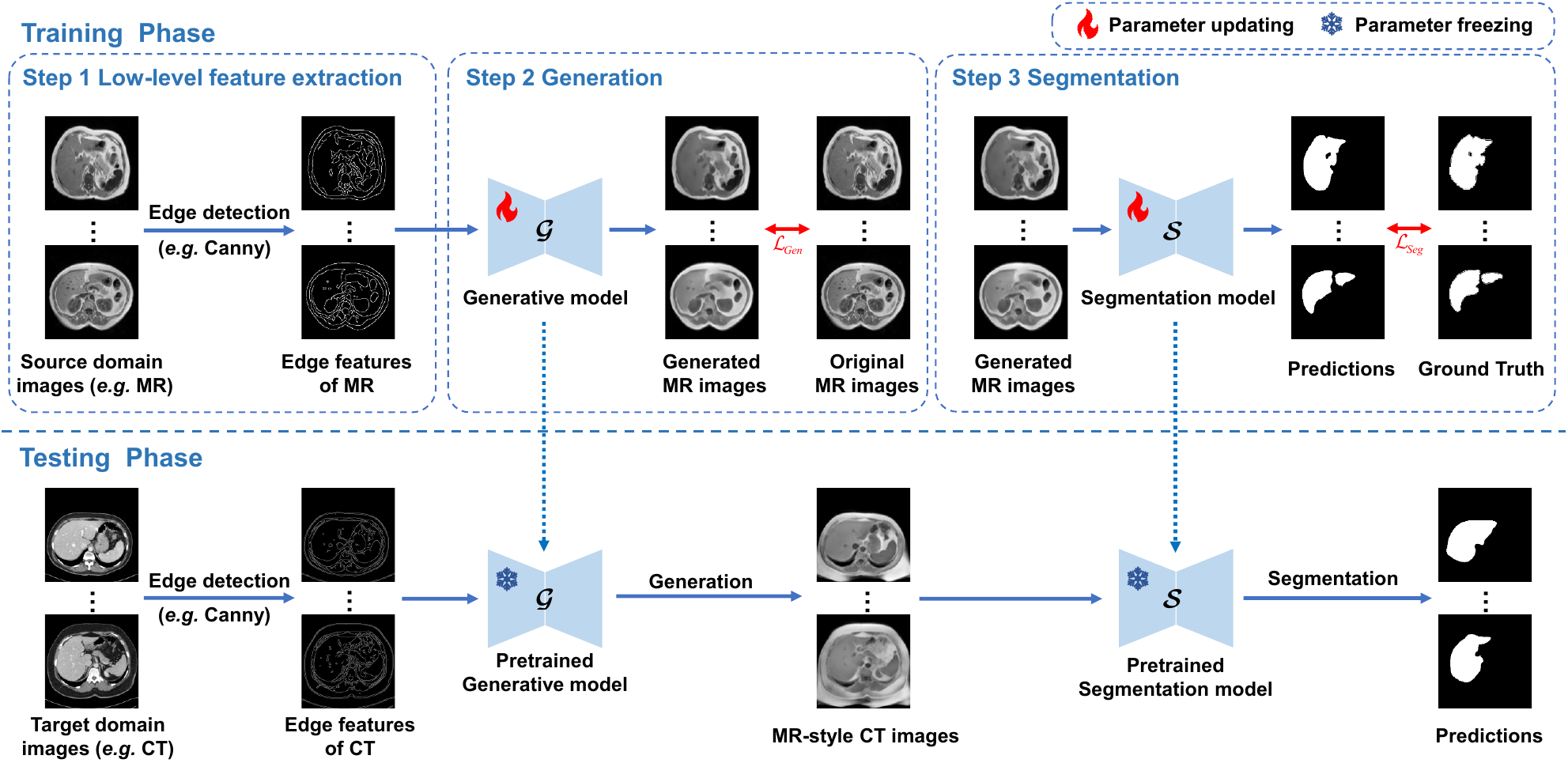}
	\caption{Framework of our \proposed. Low-level features (\eg edge features) are treated as domain-invariant representations to train a generative model, $\mathcal{G}$, to generate source-style images. Edge features extracted from unlabeled target images are then input to $\mathcal{G}$, followed by a segmentation model, $\mathcal{S}$, pretrained on the source data, to output the predictions.
    }
\label{fig:frame}
\end{figure*}

To overcome the reliance of conventional UDA on target-domain data during training,
in this paper, 
we further restrict the setting by formulating cross-modal image segmentation as a \textit{source-only} domain generalization (DG) problem,
where a model trained on labeled source data is applied to an unseen and unlabeled target modality, 
without access to any target-domain samples during training.
Effective cross-modal generalization therefore requires identifying modality-shared characteristics that are less sensitive to domain-specific variations. 
Our motivation stems from the observation that, despite significant differences in characteristics, such as intensity and contrast,
cross-modal images (\eg MRI and CT) of the same anatomical regions often preserve similar low-level structural information, particularly edges, as illustrated in Fig. \ref{fig:frame}.
In addition, edge features tend to be less sensitive to modality-specific characteristics and 
inter-individual variations,
making them potentially robust cues for cross-modal generalization.
This leads to the question: can we effectively bridge the domain gap across modalities by exploiting these similar low-level features, specifically edges? 

To answer this question, we propose \proposed, a DG framework that exploits edge information to bridge the modality gap in cross-modal medical image segmentation.
Unlike existing works \cite{cui2021structure,pei2021disentangle,gu2023cddsa} that adopt edge cues as auxiliary regularization or pursue implicit feature disentanglement, 
our method takes explicitly extracted edge maps as the core bridge for image reconstruction, 
without relying on complex feature alignment or discriminative constraints.
Our contributions are threefold:
\emph{firstly}, 
we propose \proposed with
a complete pipeline that trains a generative model to reconstruct the source images from their edge features, followed by training a segmentation model on the labeled reconstructed source images.
It requires neither test-time fine-tuning nor any complex pipelines, such as deep feature alignment and adversarial training, employed by most existing methods.
This facilitates more stable and efficient model training.
\emph{Secondly}, 
we benchmark \proposed on two publicly available MRI-CT datasets.
Despite operating in a more restrictive source-only setting (\ie without access to target-domain samples during training), 
\proposed outperforms existing state‑of‑the‑art (SOTA) UDA approaches on liver and cardiac-substructure segmentations across different settings.
\emph{Thirdly,}
we demonstrate through extensive ablation studies that \proposed can integrate with different types of generative and segmentation models, 
highlighting its model-agnostic nature and potential for further performance improvements.

\section{Method}
\label{sec:Meth}
We propose \proposed,
a framework for cross-modal medical image segmentation, as illustrated in Fig. \ref{fig:frame}. 
There is a source labeled dataset, $I^s = \left\{ {x_i^s, y_i^s} \right\}_{i = 1}^{N_s}$, 
consisting of $N_s$ pairs of labeled images and their corresponding labels,
where each image,
$x_i^s \in \mathbb{R}^{H \times W}$, 
and its label,
$y_i^s \in \mathbb{R}^{n \times H \times W}$,
have a height $H$, width $W$ and $n$ classes.
At deployment, \proposed is applied to an unseen target-domain unlabeled dataset, 
$I^t = \left\{ {x_i^t} \right\}_{i = 1}^{N_t}$,
acquired from a different modality
(\eg MRI vs. CT) but imaging the same anatomical region (\eg abdomen).
Our goal is to train a generative model, $\mathcal{G}(\cdot;\beta_{gen})$,
parameterized by $\beta_{gen}$,
and a segmentation model,
$\mathcal{S}(\cdot;\beta_{seg})$,
parameterized by $\beta_{seg}$,
using the source labeled dataset,
$I^s$.
The trained models will then be applied to the target unlabeled dataset $I^t$ for cross-modal segmentation.

\subsection{Training on the source domain}
\label{sec:training}
For source-only cross-modal DG, the objective is to learn source-domain representations that remain informative when applied to an unseen target modality.
In this work, we exploit the observation that low-level features, namely edges, in medical images are relatively invariant to domain shift,
making them a reliable foundation for 
aligning the target domain distribution with the source domain distribution for cross-modal image segmentation.
\\

\noindent \textbf{Edge feature extraction.}
We first employ an off-the-shelf edge detector $\mathcal{E}(\cdot)$ on the source domain images, $\left\{ {x_i^s} \right\}_{i = 1}^{N_s}$ to extract their corresponding edge features $\left\{ {e_i^s} \right\}_{i = 1}^{N_s}$, 
where $e^s_i \in \mathbb{R}^{H \times W}$ provides an outline of the anatomical structures present in the image. 
This process is denoted as:
\begin{equation}
\label{eq:edge}
\begin{aligned}
e^s_i = \mathcal{E}(x^s_i).
\end{aligned}
\end{equation}
Following evaluation of various edge detection algorithms, 
we selected the Canny detector \cite{canny1986computational} as our default edge detector, given its excellent noise immunity, multi-scale, and high-precision detection capabilities.
\\

\noindent \textbf{Edge-to-image reconstruction.}
Next, we train a generative model, $\mathcal{G}(\cdot;\beta_{gen})$, to reconstruct the source domain images from their corresponding edge features, $e^s_i$. 
This process can be expressed as:
\begin{equation}
\label{eq:generate}
\begin{aligned}
g^s_i = \mathcal{G}(e^s_i;\beta_{gen}),
\end{aligned}
\end{equation}
where $g^s_i\in \mathbb{R}^{H \times W}$ represents the generated source image, which aims to replicate the original source domain image $x^s_i$. 
The generative model $\mathcal{G}(\cdot;\beta_{gen})$ is trained using the mean squared error $\mathcal{L}_{MSE}$ as the loss function, denoted as:
\begin{equation}
\begin{aligned}
\mathcal{L}_{gen} = \mathcal{L}_{MSE}(g^s_i,x^s_i).
\label{eq:loss_gen}
\end{aligned}
\end{equation}

We investigate several generator architectures, including encoder-decoder architecture (\eg \cite{cao2022swin}) and adversarial architecture (\eg \cite{goodfellow2014generative}), within the proposed framework in Sec. \ref{sec:abl}.
\\

\noindent \textbf{Segmentation.}
Finally,
we train a segmentation model $\mathcal{S}(\cdot;\beta_{seg})$
using the reconstructed source images,
$\left\{ {g_i^s} \right\}_{i = 1}^{N_s}$,
and their corresponding labels,
$\left\{ {y_i^s} \right\}_{i = 1}^{N_s}$.
The segmentation predictions, 
$\left\{ {p_i^s} \right\}_{i = 1}^{N_s}$, are obtained through:
\begin{equation}
\label{eq:segment}
\begin{aligned}
p^s_i = \mathcal{S}(g^s_i;\beta_{seg}).
\end{aligned}
\end{equation}
To optimize the segmentation model,
we define the segmentation loss, 
$\mathcal{L}_{seg}$, as a combination of the cross-entropy loss,
$\mathcal{L}_{CE}$, and the Dice loss,
$\mathcal{L}_{Dice}$:
\begin{equation}
\begin{aligned}
\mathcal{L}_{Seg} = \alpha_{ce}\mathcal{L}_{CE}(p^s_i,y^s_i) + \alpha_{dice}\mathcal{L}_{Dice}(p^s_i,y^s_i),
\label{eq:loss_seg}
\end{aligned}
\end{equation}
where  
$\alpha_{ce}$ and $\alpha_{dice}$ are the weights of the respective losses.
 
In our experiments in Sec. \ref{sec:abl}, 
we instantiate the segmentation model, $\mathcal{S}(\cdot;\beta_{seg})$, with several tested architectures,
including convolutional neural networks (CNNs) \cite{ronneberger2015u}, 
Transformers \cite{cao2022swin}, 
and hybrid models. 

\subsection{Segmentation on the target domain}
At test time, we utilize the trained models from Sec. \ref{sec:training} to segment unlabeled cross-modal target domain images,
$I^t = \left\{ {x_i^t} \right\}_{i = 1}^{N_t}$. 

Specifically,
we first extract the edge feature, $e^t_i$,
from $x_i^t$ using the same edge detector $\mathcal{E}(\cdot)$ employed in Eq. (\ref{eq:edge}),
ensuring structural consistency between target and source domain edge representations.

The edge feature $e^t_i$ is then input into the pretrained generator, $\mathcal{G}(\cdot;\beta_{gen})$, to synthesize a source-style target image, $g^t_i$, as described in Eq. (\ref{eq:generate}).
By exploiting modality-shared edge information, 
$g^t_i$ inherits structural details from $x^t_i$ while adopting the appearance traits from the source domain, thereby bridging the domain gap.

Finally, the source-style target image $g^t_i$ is segmented using the trained segmentation model, $\mathcal{S}(\cdot;\beta_{seg})$, 
as outlined in Eq. (\ref{eq:segment}).
Notably, the entire inference process does not require fine-tuning on target domain data, maintaining complete test-time efficiency.

\section{Experiments and Results}
\subsection{Experimental Setup}
To evaluate the proposed \proposed, we conducted cross-modal segmentation experiments (\ie, MRI $\rightarrow$ CT and CT $\rightarrow$ MRI) for liver and cardiac substructure segmentation on two publicly available datasets. Here, $\rightarrow$ denotes knowledge transfer between modalities.

\textbf{Datasets.} For \textbf{liver segmentation,} we employed the CHAOS dataset \cite{kavur2021chaos} from ISBI 2019 CHAOS Challenge, which contains 20 sets of three-sequence MRI (\ie, T1 In-phase, T1 Out-phase, and T2) and CT images, with pixel-level liver annotations. 
The voxel spacing is $1.36 \times 1.89 \times (5.5\sim9.0) mm^3$ for MRI and $0.7 \times 0.8 \times (3.0\sim3.2) mm^3$ for CT.
For simplicity and proof of concept, we used only the T1 In-phase sequence as the MRI modality. 
For \textbf{cardiac substructure segmentation,} we utilized the MMWHS 2017 dataset \cite{zhuang2016multi}, which includes 20 unpaired MRI and CT volumetric scans, with voxel spacings of $2 \times 2 \times 2 mm^3$ and $0.44 \times 0.44 \times 0.60 mm^3$, respectively.
\begin{table}[!tbp]
    \fontsize{9}{9}\selectfont
    \caption{Quantitative comparison between our \proposed and existing SOTA models on the CHAOS dataset for liver segmentation. Best results are in \textbf{bold}. “-” Indicates the data is not available.}
    \centering
    \begin{tabular}{c|c|c||c|c}
        \toprule[1.0pt]
        & \multicolumn{2}{c||}{Liver MR $\rightarrow$ CT} & \multicolumn{2}{c}{Liver CT $\rightarrow$ MR}\\	
        \hline
        Methods & Dice $\uparrow$& ASD$\downarrow$ & Dice $\uparrow$ & ASD$\downarrow$\\
        \hline
        Supervised training &94.3 &1.0 &91.9 &0.7\\
        W/o adaptation &50.5 &25.4 &42.2 &17.1\\
        \hline
        SIFAv1$_{19}$ \cite{chen2019synergistic} &78.7 &9.6 &66.5 &12.3\\
        SIFAv2$_{20}$ \cite{chen2020unsupervised} &79.2 &9.4 &67.8 &11.6 \\
        DDFseg$_{21}$ \cite{pei2021disentangle} & 78.4 &10.2 &-- &--\\
        MPSCL$_{22}$ \cite{liu2022margin} &81.8 &4.9 &75.2 &5.6\\
        M$^{2}$CD$_{23}$ \cite{pei2023multi} &80.3 &6.5 &-- &--\\
        SIAB$_{25}$ \cite{qiu2025devil} &75.6 &6.8 &72.1 &9.8\\
        TCSA$_{26}$ \cite{TCSA-UDA} & 79.5 & 6.2 & 73.6 & 6.8\\
        \hline
        Ours‑LowBridge & \textbf{84.6} &\textbf{4.5} &\textbf{78.3} &\textbf{4.8} \\
        \bottomrule[1.0pt]
    \end{tabular}
    \label{table:liver_seg}
\end{table}
\begin{figure}[!htp]
	\centering \includegraphics[width=0.48\textwidth]{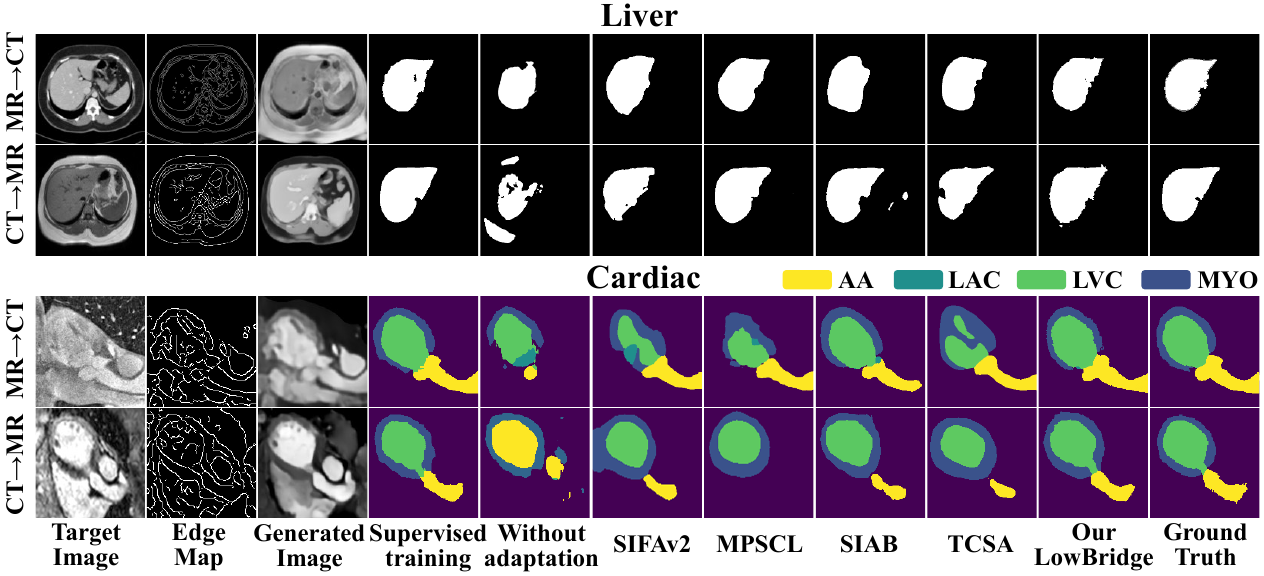}
	\caption{Visual results on the CHAOS and MMWHS datasets.}
\label{fig:comp}
\end{figure}
Both modalities have pixel-wise annotations of four cardiac structures: ascending aorta (AA), left atrium blood cavity (LAC), left ventricle blood cavity (LVC), and myocardium of the left ventricle (MYO). For a fair comparison, we processed and split the data following \cite{chen2020unsupervised}.

Under the proposed source-only DG setting, neither target-domain images nor labels were used during training, with target labels accessed only for final evaluation. Unless otherwise stated, all images were processed slice-wise in 2D, while 3D UNet ablation used volumetric inputs under the same source-target split protocol.

\textbf{Implementation Details.} 
For the generative model $\mathcal{G}(\cdot;\beta_{gen})$,
we used FreUNet, a UNet variant equipped with a pretrained CDFFormer-M \cite{tatsunami2024fft} encoder.
For the segmentation models $\mathcal{S}(\cdot;\beta_{seg})$, we employed 
SwinUNet \cite{cao2022swin}.
The loss weights ($\alpha_{ce}$ and $\alpha_{dice}$) in Eq. (\ref{eq:loss_seg}) were set to 1.
All models were implemented in PyTorch and trained for 100 epochs on an NVIDIA Tesla P100 GPU.
FreUNet and SwinUNet were optimized using Adam and AdamW,
respectively, with batch sizes of 6 and 4
and a learning rate of $1\times10^{-4}$ for both.
Data augmentations like random cropping, multi-angle rotation, and intensity enhancement were applied to prevent overfitting.

\textbf{Evaluation Metrics.} 
We adopted Dice and Average Surface Distance (ASD) to evaluate segmentation accuracy, and PSNR to assess generation performance.
Better performance is indicated by higher ($\uparrow$) Dice and PSNR values and lower ($\downarrow$) ASD values.

\begin{table}[ht]
\centering
\fontsize{9}{9}\selectfont
\caption{Quantitative comparison between our \proposed and existing SOTA models on the MMWHS dataset for cardiac substructure segmentation. Best results are in \textbf{bold}.}
\setlength{\tabcolsep}{2pt} 
\begin{tabular}{c|c|c|c|c|c||c}
\toprule[1.0pt]
& \multicolumn{6}{c}{Cardiac MR $\rightarrow$ CT}\\	
\hline
\multirow{2}{*}{Methods} & \multicolumn{5}{c||}{Dice $\uparrow$}& ASD$\downarrow$\\
\cline{2-7}
& AA & LAC & LVC & MYO & Average & Average\\
\hline
Supervised training &95.1 &93.0 &90.1 &89.4 &91.9 & 1.6\\
W/o adaptation &49.3 &57.3 &59.3 &63.7 &57.4 &12.8\\
\hline
SIFAv1$_{19}$ \cite{chen2019synergistic} &81.1 &76.4 &75.70 &58.7 &73.0 &8.1\\
SIFAv2$_{20}$ \cite{chen2020unsupervised} &81.3 &79.5 &73.8 &61.6 &74.1 &7.0\\
DSAN$_{21}$ \cite{han2021deep} &79.9 &84.8 &82.8 &66.5 &78.5 &5.9\\
MPSCL$_{22}$ \cite{liu2022margin} &90.3 & 87.1 & 86.5 &72.5 &84.1 &3.5\\
SE-ASA$_{23}$ \cite{feng2023unsupervised} &83.8 &85.2 &82.9 &71.7 &80.9 &5.4\\
MA-UDA$_{24}$ \cite{ji2023unsupervised} &90.8 &88.7 &77.6 &67.4 &81.1 &5.6\\
SIAB$_{25}$ \cite{qiu2025devil} &88.8 & 83.8 &\textbf{89.1} &77.1 &84.7 &5.7\\
TCSA$_{26}$ \cite{TCSA-UDA} & 82.5 & 87.1 & 85.7 & 74.3 & 82.4 & 6.9\\
\hline
Ours-LowBridge &\textbf{94.6} &\textbf{92.0} & 84.6 & \textbf{78.4} & \textbf{87.4} &\textbf{2.8}\\
\midrule[1.0pt] 
& \multicolumn{6}{c}{Cardiac CT $\rightarrow$ MR}\\		
\hline
\multirow{2}{*}{Methods} & \multicolumn{5}{c||}{Dice $\uparrow$}& ASD$\downarrow$\\
\cline{2-7}
& AA & LAC & LVC & MYO & Average & Average\\
\hline
Supervised training &86.4 &83.5 &92.6 &82.1 &86.2 & 2.6\\
W/o adaptation &43.4 &8.9 &55.1 & 24.6 & 33.0 &17.7\\
\hline
SIFAv1$_{19}$ \cite{chen2019synergistic} &67.0 &60.7 &75.1 &45.8 &62.1 &6.2\\
SIFAv2$_{20}$ \cite{chen2020unsupervised} & 65.3 &62.3 &78.9 &47.3 &63.4 &5.7\\
DSAN$_{21}$ \cite{han2021deep} & 71.3 &66.2 &76.2 &52.1 &66.5 &5.4\\
MPSCL$_{22}$ \cite{liu2022margin} & 64.7 & \textbf{77.3} & 81.6 &55.9 &69.9 & 3.8\\
SE-ASA$_{23}$ \cite{feng2023unsupervised} &68.3 &74.6 &81.0 &55.9 &69.9 &4.3\\
MA-UDA$_{24}$ \cite{ji2023unsupervised} &71.0 &67.4 &77.5 &59.1 &68.7 &5.3\\
SIAB$_{25}$ \cite{qiu2025devil} &56.8 &50.1 &69.9 &59.8 &59.2 &7.9 \\
TCSA$_{26}$ \cite{TCSA-UDA} &69.0 &74.9 & 83.5& 59.2& 71.6&5.1\\
\hline
Ours-LowBridge & \textbf{71.7} & 69.9 & \textbf{89.0} & \textbf{65.9} & \textbf{74.1} &\textbf{3.6}\\
\bottomrule[1.0pt]
\end{tabular}
\label{table:cardiac_seg_MR2CT}
\end{table}

\subsection{Model Comparison}
We compared \proposed with ten SOTA methods, including 
(1) one medical DG method, SIAB \cite{qiu2025devil}, and
(2) nine medical UDA methods, namely SIFAv1 \cite{chen2019synergistic}, SIFAv2 \cite{chen2020unsupervised}, 
DSAN \cite{han2021deep}, DDFseg \cite{pei2021disentangle}, MPSCL \cite{liu2022margin}, M$^{2}$CD \cite{pei2023multi}, SE-ASA \cite{feng2023unsupervised}, MA-UDA \cite{ji2023unsupervised}, and TCSA \cite{TCSA-UDA}.
Our \proposed operates in a
more restrictive source-only setting (\ie without access to target-domain samples during training) than the UDA baselines (\ie with access to target-domain images during training).
Therefore, the comparison provides a challenging reference setting for \proposed.
The results of each method were obtained from the corresponding paper or running the released code.

{\textbf{Liver Segmentation.}} As shown in Table \ref{table:liver_seg} and Fig. \ref{fig:comp}, our \proposed outperformed ten existing SOTA methods. 
Compared with the top-performing MPSCL \cite{liu2022margin}, \proposed achieved improvements of 3.8$\%$ in Dice and 11.4$\%$ in ASD, respectively.

{\textbf{Cardiac Substructure Segmentation.}} 
Table \ref{table:cardiac_seg_MR2CT} and Fig. \ref{fig:comp} demonstrate the SOTA performance of \proposed.
Specifically, \proposed surpassed MPSCL \cite{liu2022margin} by 4.9$\%$ in average Dice and 12.3$\%$ in average ASD. 
Although slightly lower performance was observed for some structures,
such as LVC in MR $\rightarrow$ CT and LAC in CT $\rightarrow$ MR,
\proposed still outperformed existing UDA methods overall.
These performance drops may reflect limitations of hand-crafted edge extraction
in low-contrast or heterogeneous regions,
which should be addressed in future work.

Notably, \proposed outperforms UDA baselines despite having no access
to target-domain samples during training.
This highlights \proposed's potential for clinical applications, where target-modality data may not always be available before deployment.

\subsection{Ablation Studies}
\label{sec:abl}
To assess the architectural sensitivity of \proposed, 
we evaluated various generative and segmentation models on the CHAOS dataset. 

\begin{table}[!tbp]
    \fontsize{9}{10}\selectfont
    \caption{Ablation Studies of \proposed on the CHAOS Dataset.}
    \setlength{\tabcolsep}{0.7pt} 
    \centering
        \begin{tabular}{c|c|c|c|c|c|c|c|c}
            \toprule[1.0pt]
            &&& \multicolumn{3}{c|}{Liver MR $\rightarrow$ CT} & \multicolumn{3}{c}{Liver CT $\rightarrow$ MR}\\	
            \hline
            No. & Gen. & Seg. & PSNR$\uparrow$ & Dice$\uparrow$& ASD$\downarrow$ & PSNR & Dice$\uparrow$ & ASD$\downarrow$\\
            \hline
            1&FreUNet & SwinUNet& \textbf{21.69} &84.6 & \textbf{4.5} & \textbf{27.65} &\textbf{78.3} & 4.8 \\
            2&FreUNet & FreUNet & \textbf{21.69} &86.1 &5.0 & \textbf{27.65} &76.1 &8.6 \\
            3&FreUNet & 3D UNet & \textbf{21.69} & 82.8 & 5.6 & \textbf{27.65} & 77.8 & \textbf{4.3} \\
            4&FreUNet & UNet & \textbf{21.69} &\textbf{86.2} & 5.7 & \textbf{27.65} & 76.6 & 9.4 \\
            \hline
            5&GAN& UNet & 20.15 &76.8 &10.7 & 17.02 &59.4 &7.3 \\
            6&DDPM & UNet & 17.44 & 70.7 & 12.1 & 20.33 & 69.4 & 10.1 \\
            7&SwinUNet & UNet &19.14 &80.4 &6.2 & 25.74 &73.6 &6.9 \\
            \hline
            8& Edge & UNet & - & 64.5 & 15.4 & - & 58.2 & 13.7 \\ 
            \bottomrule[1.0pt]
        \end{tabular}
    \label{table:abl}
\end{table}

\textbf{Segmentation models}. We evaluated CNN-based UNet \cite{ronneberger2015u}, 3D UNet \cite{3DUNet}, SwinUNet \cite{cao2022swin}, and hybrid-model FreUNet,
while fixing FreUNet as the generative model. 
\begin{figure}[!htp]
	\centering \includegraphics[width=0.48\textwidth]{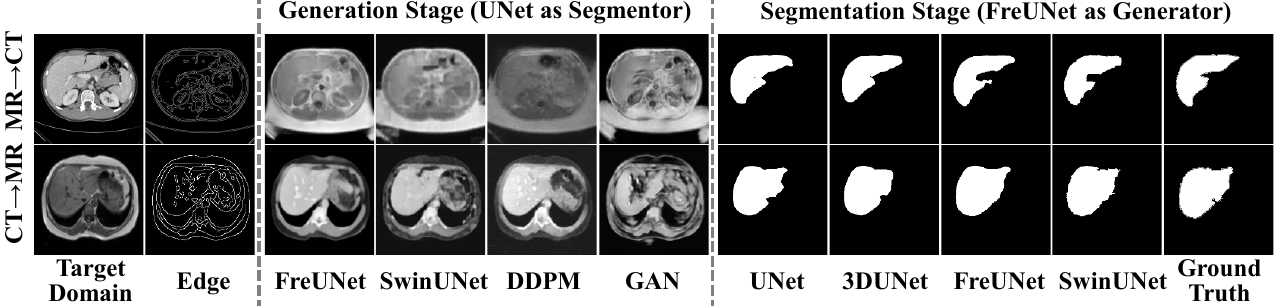}
	\caption{Visual results of the ablation studies on the CHAOS dataset.}
\label{fig:abl}
\end{figure}
As shown in Table \ref{table:abl}, the four segmenters achieved
comparable performance, with average absolute deviations of 0.38 in Dice and 1.19 in ASD. 
Despite these variations, they outperformed most existing methods in Table \ref{table:liver_seg}. 

\textbf{Generative models}. We investigated different generative models for \proposed,
including GAN \cite{goodfellow2014generative} (with a pretrained ResNet-50 \cite{he2016deep} encoder), DDPM \cite{DDPM}, SwinUNet \cite{cao2022swin}, and FreUNet, with UNet \cite{ronneberger2015u} as the segmentation model.
As shown in Table \ref{table:abl}, 
the four generators yielded varying performance, with an average PSNR gap of 6.08 between the best and worst models. Overall, higher generation quality was associated with better final performance.

\textbf{Edge-only segmentation.}
Training the segmenter directly on edge maps (row 8, Table \ref{table:abl}) yielded substantially lower performance, highlighting the importance of image reconstruction for \proposed.

\textbf{Inference speed.}
On an NVIDIA Tesla P100 GPU, the standalone generative (\ie FreUNet) and segmentation (\ie SwinUNet) models process 43.6 and 67.1 slices per second, respectively, with negligible Canny latency. Despite multi‑stage sequential overhead, \proposed maintains practical inference efficiency for clinical use.

\textbf{Limitations and future work.}
Our ablation studies demonstrate that \proposed is compatible with various tested generative and segmentation architectures.
While different segmenters achieved comparable or better performance than most existing methods in Table \ref{table:liver_seg}, the relatively poor performance of GAN and DDPM suggests that the primary bottleneck of \proposed's performance 
lies in the capability of the generative model to reconstruct high-quality images from their low-level edge features.
Moreover, this work primarily exploits hand‑crafted Canny edges as low-level features for MRI-CT segmentation.
Future endeavour may continue the investigation on the robustness of other low-level features for more general cross-modality medical image segmentation.

\section{Conclusion}
In this paper, we proposed \proposed, a source-only DG framework for cross-modal medical image segmentation. 
By treating low-level edge features as domain-invariant representations,
we designed a simple yet effective strategy to reconstruct source-style target-domain images for segmentation. 
Extensive experiments on two publicly available MRI-CT benchmark datasets demonstrated that \proposed achieves SOTA performance, surpassing ten existing methods without access to target-domain samples during training. 
Moreover, ablation studies confirmed its flexibility and compatibility with various generative and segmentation models, highlighting its potential for further improvement with more advanced techniques. This work opens up new possibilities for effective cross-modal generalization in medical image segmentation, paving the way for more robust and generalizable solutions in clinical applications.

\vfill\pagebreak

\section{Acknowledgements}
This work was supported in part by the fund of Supporting the Reform and Development of Local Universities (Disciplinary Construction) and the special research project of First-class Discipline of Inner Mongolia A. R. of China under Grant YLXKZX-ND-036, in part by the Funds from Youth Project of Liaoning Provincial Department of Education of China (grant number LJ212510157013), and in part by the China Scholarship Fund.

\bibliographystyle{IEEEbib}
\bibliography{ref}

\end{document}